\documentclass[review,numbers,sort&compress]{elsarticle}
\usepackage{natbib}
\usepackage{lineno,hyperref}
\usepackage{graphicx}
\usepackage{subfigure}
\usepackage{tikz}
\usepackage{pgfplots}
\usepackage{amsmath,bm}
\usepackage{multirow}
\usepackage{extarrows}
\usepackage{amsfonts}
\usepackage{mathrsfs}

\usepackage{booktabs}
\usepackage{algpseudocode}
\usepackage{algorithm}
\usepackage{algorithmicx}
\usepackage{arydshln} 
\usepackage{xcolor, soul}
\usepackage{ulem}
\sethlcolor{blue}
\soulregister\cite7
\soulregister\ref7
\soulregister\textbf7
\modulolinenumbers[5]

\usepackage{changes}

\journal{Information Sciences}
\begin{document}
\begin{frontmatter}

\title{Video and Audio are Images: A Cross-Modal Mixer for Original Data on Video-Audio Retrieval}

\author{Zichen Yuan}
\author{Qi Shen}
\author{Bingyi Zheng}
\author{Yuting Liu\corref{cor}}
\cortext[cor]{Corresponding authors}
\author{Linying Jiang}
\author{Guibing Guo\corref{cor}}

\address{Northeastern University, China}

\begin{abstract}
Cross-modal retrieval has become popular in recent years, particularly with the rise of multimedia. Generally, the information from each modality exhibits distinct representations and semantic information, which makes feature tends to be in separate latent spaces encoded with dual-tower architecture and makes it difficult to establish semantic relationships between modalities, resulting in poor retrieval performance. To address this issue, we propose a novel framework for cross-modal retrieval which consists of a cross-modal mixer, a masked autoencoder for pre-training, and a cross-modal retriever for downstream tasks. In specific, we first adopt cross-modal mixer and mask modeling to fuse the original modality and eliminate redundancy. Then, an encoder-decoder architecture is applied to achieve a fuse-then-separate task in the pre-training phase. We feed masked fused representations into the encoder and reconstruct them with the decoder, ultimately separating the original data of two modalities. In downstream tasks, we use the pre-trained encoder to build the cross-modal retrieval method. Extensive experiments on 2 real-world datasets show that our approach outperforms previous state-of-the-art methods in video-audio matching tasks, improving retrieval accuracy by up to $2\times$. Furthermore, we prove our model performance by transferring it to other downstream tasks as a universal model. 
\end{abstract}


\end{frontmatter}

\section{Introduction}

\begin{figure*}[t]
\centerline{\includegraphics[width=1\textwidth,height=0.3\textwidth]{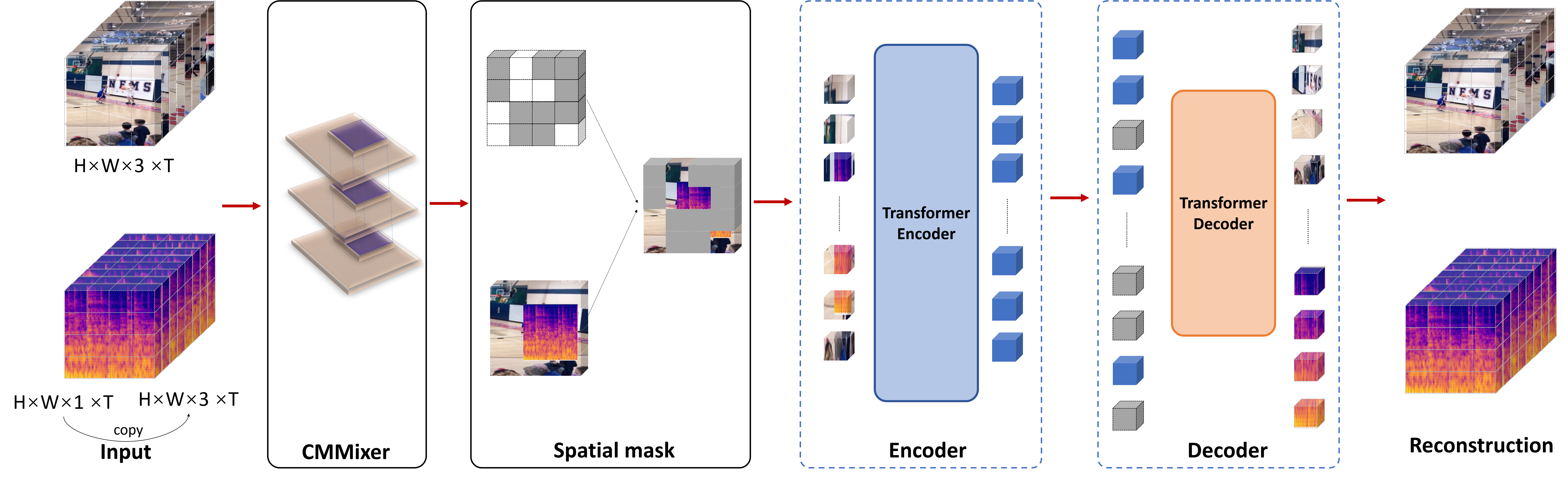}}
\caption{Pre-training with CMMixer: We extract audio and video recordings and then embed the video into image space while converting the audio spectrogram by similar image space by copying it into three channels. Next, we use our proposed Cross-Modal Mixer (CMMixer) technique to fuse the modalities. We then mask out a large subset of random patches and use a subset of masked tokens. Combined with the shared visible encoder patches, we use an encoder to encode the cross-modal information and reconstruct the original date of two modalities using a small decoder.}
\label{fig:main}
\vspace{-0.38mm}
\end{figure*}

The rise of micro-videos has led to an increased interest in matching music to videos. A tool that assists creators in finding the ideal music or matching video footage to some specific songs can be incredibly useful and provide new creative possibilities. For example, Creators in Tiktok~\footnote{\footnotesize \url{https://www.tiktok.com/}} could use the tool to find trending songs or music that matches the vibe of their videos. This helps them jump on trends and create content that resonates with viewers. However, scoring music for micro-videos is a comprehensive work that requires balancing between subjectivity and objectivity, rationality and sensibility, as well as individuality and generality. It requires a high level of musical accomplishment, aesthetic ability and skill, which leads to a result that matching music to micro-videos can be quite challenging work.


In practice, we can divide this task into two categories based on the patterns of interaction between factors. This is saying that the different elements that compose and influence the micro-video can be considered as factors, such as the music, choreography, camera work, editing, etc. Each of them may have a latent or manifest relationship with the other, which we define this pattern as interaction. Analyzing these interactions between factors is key to understanding what makes a quality micro-video. The first category includes straightforward examples where an appropriate audio-video pair can be easily matched due to obvious common factors, like connecting a Christmas scenario to the song "Silver Bells". However, this kind of video-audio pair with pure, discrete factors is relatively uncommon. More often, finding suitable music for a complex video like K-pop dancing is challenging since the relevant factors are nuanced and entangled. This second, more realistic category where latent interactions predominate represents the majority of real-world examples. Though matching music and video in this category is difficult, continuing to study these complex cases is valuable, as it will enable more robust and widely applicable solutions. Therefore, to achieve satisfactory performance in matching audio and video, we must focus on handling hard examples which key is to capturing the relationship between modalities, which is bound by some particular factors.

Unfortunately, there are only a few researchers made efforts to the corresponding work about matching video and audio, and previous methods for matching them are not fully well-studied. Those methods relied on uniformly splitting the video and audio into segments of equal duration \cite{pretet2021cross, yi2021cross, suris2022s}, serializing the extracted fragment or frame of segments as sequences and feeding them into the model orderly to get embeddings without context interactive, using them to predict the degree of matching directly. However, to grasp the relationship between a video and its music, focusing on a single segment ignores continuous echo and progression between preceding and following segments. This makes it difficult to achieve congruence in emotion and rhythm between the video content and the music. Thus, they can not capture long-term temporal artistic correspondence efficiently.  Specifically, the method of MVPt~\cite{suris2022s} fails to address the challenges of hard examples as it does not effectively model the temporal dynamics and interactions between video and audio. Their reliance on pre-trained vision features~\cite{chen2020simple} and audio features alone is insufficient. Therefore, it only has an average performance in tasks with easy examples, while has a worse performance in hard ones. In response to the aforementioned problems, we need to optimize the process of handling the multimodal information by considering their relations and building a framework to model the interdependence of video and audio information while taking in the long-range temporal context.

To address these limitations, we propose a new framework that optimizes multimodal fusion to model the interdependencies between video and audio. This allows capturing long-range temporal context critical for handling hard examples where factors are entangled. Recently, various efforts~\cite{cheng2022vista,wang2022vlmixer,chen2022improving,wang2022eclip} have been put into designing new fusion paradigms that can effectively fuse multimodal data which are generally task-specific. For audio and video, the topology of their information representations is different since the time-frequency representations have distinguishable distributions. The visual stream in a video is three-dimensional (two spatial and one temporal), while the time-frequency stream in audio is two-dimensional (one frequency and one temporal). Thus, we need to design specialized input representations to accomplish specific tasks.

Moreover, we observe that both the visual frames of the video and the spectrogram audio clip have redundancy because the matching process mostly relies on particular entangled factors and overall context. Therefore, we draw lessons from the successes of BERT~\cite{devlin2018bert} in NLP and MAE~\cite{he2022masked} in CV, innovating an effective pre-processing method of multimodal data and leveraging a mixer to emphasize the frame-level synchronization. Specifically, we first extract video and audio data ordered by time stamps, mapping both images and mel-spectrums into three channels, obtaining a three-channel representation through duplicating the mel-spectrums inspired by the multi-head attention. With the additional dimension, the model can implicitly perform multi-source separation during encoding which makes it possible to disentangle audio from different objects at the same time. Second, because of the heavy spatial redundancy in images~\cite{he2022masked}, we randomly drop part of the data to improve encode efficiency after mixing them to gain a mixture. Finally, we design a masked modeling framework to take in all temporal mixed representations in one encoder and reconstruct modalities separately in pre-training. Note that the framework can optionally accept additional modalities and explicitly learn modality-common representations. In addition, we use contrastive learning to distinguish between positive and negative pairs based on the inherent semantics of cross-modal data to handle hard examples.

Compared to the previous methods, our pre-training strategy utilizes mixed long-range cross-modal temporal information and captures the correlation between modalities implicitly without redundancy. We can use it as an intermedia to simplify pairing audio and video without interaction by enabling access to modality semantic information, which helps us to handle the hard examples successfully.

To sum up, we make the following contributions.
\begin{itemize}
\item  We propose a novel cross-modal fusion method called \underline{C}ross-\underline{M}odal \underline{Mixer} (CMMixer). It transforms mel-spectrums into image-like representations based on multi-source separation and conducts cross-modal fusion on the raw data to provide temporal mixed information. This converts time-frequency representations into image-like distributions.

\item We develop a new multimodal pre-training framework based on Masked Autoencoder (MAE) that uses a single encoder-decoder structure. Our model employs a single modal structure to capture cross-modal dependencies and common representations implicitly via a fuse-then-separate strategy with attention. This pre-training framework enhances model expressivity. We fine-tune this framework for audio-video retrieval.

\item To evaluate the effectiveness of our proposed method, we created and publicly released M2M dataset. We conduct experiments on both public datasets and our dataset for evaluation. The experimental results show that our model outperforms state-of-the-art retrieval methods.
\end{itemize}

The remainder of this paper is organized as follows: Section 2 reviews related works on visual and video representation learning and cross-modal matching of music and video. Section 3 presents our proposed approach, CMMixer. Section 4 experimentally evaluates CMMixr. Section 5 analyses the ability of our model using visualization of the cluster of embedding and reconstruction result and explore a latent novel application scenario of it. Section 6 concludes the paper and discusses directions for future work.

\section{Related Work}
\subsection{Visual and Video Representation Learning}

Since the introduction of Transformers \cite{vaswani2017attention}, self-supervised modeling has made substantial progress in recent years. After pre-training on a large amount of unlabeled data with reconstruction tasks, the success of masked language models in natural language processing, such as BERT, has enlightened methods for self-supervised learning. Taking inspiration from the success of NLP, much research focuses on implementing similar methods on CV which involve using different proxy tasks during pre-training to improve performance. The autoencoder network presented in BEiT \cite{bao2021beit}, SimMIM \cite{xie2022simmim} and MAE are methods for the masked image modeling (MIM) task in CV. They explore different methods and modal architecture to find effective ways for image-masked autoencoders. MixMIM \cite{liu2022mixmim} and i-MAE \cite{zhang2022mae} studied the separability and the degree of semantics on the latent features of the mixed images. 

For long-form information, such as video, self-supervised learning often aims to make use of the temporal dimension in videos. For example, modeling object motion \cite{agrawal2015learning,wang2015unsupervised,pathak2017learning,wang2019learning} and temporal ordering \cite{misra2016shuffle,fernando2017self,lee2017unsupervised,wei2018learning,xu2019self} or predicting the future \cite{walker2016uncertain,vondrick2016anticipating,mathieu2015deep,lotter2016deep,vondrick2018tracking,diba2019dynamonet}. Approach based on the masked modeling is to use a high masking ratio and reconstruction, as videos and audio tend to be more redundant in terms of information \cite{huang2022masked, tong2022videomae, feichtenhofer2022masked}. Further research, M3AE \cite{geng2022multimodal}, beitv2 \cite{peng2022beit} and v3 \cite{wang2022image} try to improve training efficiency and get more generalizable representations by introducing information from different modalities to provide rich supervision.

Besides reconstruction, contrastive learning is another significant approach that focuses on modeling the similarities and differences between multiple views of images or texts \cite{xie2022simmim,he2020momentum,oord2018representation,grill2020bootstrap,wu2018unsupervised,radford2021learning,jia2021scaling}. Some examples of methods using contrastive learning include SimCSE \cite{gao2021simcse}, which creates positive pairs of sentences through the application of Dropout, and SimCLR \cite{chen2020simple}, which uses random image augmentation. However, contrastive learning often relies heavily on data augmentation, which can potentially introduce bias during the training process.

In this work, our proposed Cross-Modal Mixer (CMM) module, a multimodal self-supervised fusion algorithm is an early fusion approach to produce a cross-modal view to adapt the single-masked autodecoder architecture to get a better modality-common representation.

\subsection{Cross-modal Matching of Music and Video}
There have been numerous efforts to develop frameworks for recommending music to accompany a given video. However, many of these approaches have limitations. Some only consider the mood of the video and the user's listening history \cite{kuo2013background,shah2014advisor}. These methods require manual annotations for each video and audio segment, which is high in labor and time costs, and are limited to a small number of predetermined mood categories \cite{yi2021cross}. Others \cite{shah2014advisor,li2019query,zeng2018audio} use cross-modal ranking losses to match signals between music and videos without any metadata like emotions. Further work employed the visual features and audio features provided by pre-trained feature extraction models \cite{suris2022s} to constrain the visual and audio embeddings of the same video as closely as possible. However, these approaches still have limitations, such as the need for manual annotations or the excessive reliance \cite{pretet2021cross,yi2021cross} on certain metadata of the video. 

There have been attempts to generate MIDI files based on finger movements \cite{gan2020foley,su2020audeo,di2021video}, but these approaches are limited in capturing low-level signals and achieving stability over time. Research \cite{suris2022s} has explored the connection between music traits and video editing operations through interviews with professional editors and analysis of existing video data. Cross-modal retrieval research \cite{wang2022vlmixer,chen2022improving,cheng2022vista}, in which scene and text information is used, has developed an efficient method by using implicit cross-modal alignment learning to enhance cross-modal understanding. 

Our method indicates that a masked autoencoder with temporal context only can be sufficient in establishing good modality-common representation between audio and video data without other constraints. Then, our work builds upon previous efforts by using a contrastive learning framework with InfoNCE contrastive loss \cite{oord2018representation} to fine-tune the modality-common representation in the matching phase and significantly improve retrieval accuracy.

\section{Our Method}

In this section, we introduce the proposed modules, including \textbf{Cross-Modal Mixer}, \textbf{Masked Autoencoder}, and \textbf{Multimodal Retrieval Network}. The whole framework consists of an encoder and decoders which are based on ViT-Base and a two-stream retrieval model. And we also create a public dataset \textbf{M2M} and conduct experiments on it. The overall pipeline of our framework is illustrated in Fig.~\ref{fig:main}.

\subsection{Introduction of The M2M dataset}
We have created M2M, a balanced dataset that addresses the lack of publicly available resources that pair videos with background music. It is intended to serve as a benchmark for micro-video background music recommendations. We collected a set of short videos from TikTok by uniformly searching according to category labels. The dataset includes both manually paired audio clips and videos, as well as videos with audio collected from its environments. This dataset allows the model to better learn the similarity between video and audio, and train to recognize pairs in real-world scenarios. The primary goal of M2M is to accurately reflect the distribution of micro-videos and candidate audio clips in real-world scenarios, in order to facilitate research on the relationship between background music and micro-videos. Unlike other datasets such as Youtube-8m \cite{abu2016youtube} and Kinetics \cite{carreira2017quo}, which mainly associate a jumbled background sound to a video, M2M closely resembles the way music and videos appear in the real world, providing a more meaningful and useful resource for researchers.

\subsection{Cross-Modal Mixer}
\begin{figure}[htbp]
    \centering
    \includegraphics[scale=0.578]{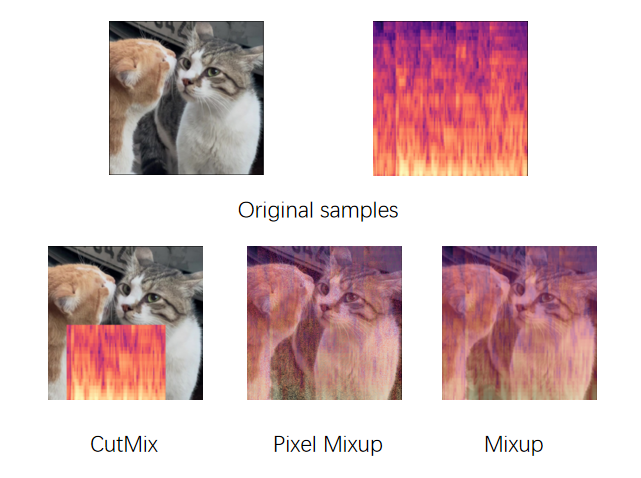}
    \caption{\textbf{Overview of the Channel Combination}. We use CutMix, Pixel Mixup, and Mixup to combine two modalities. However, due to the challenges in displaying accurate visual information of the spectrogram with three channels, the mixed image visualization is only a schematic representation.
}
    \label{fig:mcc}
\end{figure}

We expand some methods in images-only and text-only to cross-modal modeling, including an image-fusing model and a sequential encoding model. First, a video of length $T$ can be partitioned into $K$ clips as $K=T/t$, each of length $t$. By sampling the video and audio content at the temporal midpoint of each clip, we obtain two sets of information, $I^v = \{x_{1}^v,x_{2}^v,...,x_{K}^v\}$ and $I^a = \{x_{1}^a,x_{2}^a,...,x_{K}^a\}$. Then we obtain the mid-frame of the processed video segment and its corresponding Mel-spectrogram. During the training, we aim to create a new training sample $I^i$, the mixture of video $I^v$ and audio $I^a$ through unsupervised learning, and utilize a special array mode to keep their time sequence strictly. The universal fusing method can be formulated as Eq.\ref{eq1}:

\begin{equation}
I^i=M\odot I^v+(1-M) \odot I^a, \label{eq1}
\end{equation}
where $\odot$ is element-wise multiplication, $M$ is a coefficient matrix, and the key to fusion lies in the calculation of $M$. To be specific, we describe several lateral instantiations here. Our method operates on pixels in 3 types(Shown in Fig.~\ref{fig:mcc}) and we verify the most effective way among them. 

Here, for the fusion between the two types of data, we describe three methods in detail below. The image is defined as the size of $H$ by $W$ corresponding to the height and width. The position of each pixel in an image can be represented by a two-dimensional coordinate in the form of $(x,y)$. The coefficient matrix $M$ is a single-channel matrix of size $H$ by $W$ and has elements of only 1 and 0, using the element-wise multiplication with an image to represent the selection and non-selection of a pixel of this position respectively. If an element has a value between 0 and 1, it indicates the partial selection of that pixel. 

\begin{itemize}
    \item Cutmix: The process is based on CutMix, which involves randomly selecting location bounding box $B_i=(r_x,r_y,r_w,r_h)$, where $r_i$ is a two-dimensional coordinate. The technique can generate set of $s$ boxes $B^s=\{B_1,B_2...,B_n\}$ without overlap. By combining $B^s$, we determine the regions used for data fusion in the $H\times W$-sized image. For the coefficient matrix $M$, the elements bounded by the box were assigned the value 1; outside the box, the elements were assigned the value 0.
   \item Pixel Mixup: The Pixel Mixup adopts another method to generate the coefficient matrix. In particular, we define a vector to achieve this function. This vector can be represented as $V=(v_1,v_2,...,v_{H \times W})$,  Here, .$v_i$ is randomly assigned to 1 with probability $p$, otherwise it is 0. And then we reshape it into the coefficient matrix $M$ of size $H$ by $W$. 
    \item Mixup: We define the combination ratio $\lambda$  sampled from the beta distribution $\beta$($\alpha$,$\alpha$). The coefficient matrix $M$ also has the size of $H$ by $W$ and the value of each element of $M$ is $\lambda$.
\end{itemize}

\subsection{CMMixer with Masked Autoencoder}
The MAE architecture in the “mask-then-regenerate" training method consists of two networks: an encoder and a decoder. Following the architectures of ViT, BERT and MAE, the encoder consists of piles of Transformer blocks and takes only visible mixture tokens $I^r$ processed by CMMixer as input. Then, the patches are embedded by linear projection, and added with position embedding. We use a variant of position embedding known as the 2D sine-cosine method \cite{devlin2018bert}, which is more suited for handling more complex and multi-modal sequential information instead of the traditional learnable 1D position embedding in the original Vision Transformer (ViT) model. 

In the decoder, we have a reconstruction task for two different modalities: video and audio. To handle this, we use a single decoder that takes the shared representational space of the two modalities'  full set of tokens $I^f$ consisting of mask tokens $m$ and encoded visible tokens $p$ as input, $I^f = \{m,p\}$. Each mask token is an embedding vector that represents a missing patch that needs to be predicted by its corresponding decoder $D$. The input for the decoder is composed of the mask token and visible tokens. 

Additionally, the decoder also uses the same 2D position embeddings. The decoder consists of other stacks of Transformer blocks, and its final layer is composed of shared linear projections that output the same number of channels as the number of pixel values in a patch, which represent the reconstruction $R^v$, $R^a = D(I^f)$ of video and audio information respectively. 

In addition, note that the decoders are only used during the pre-training phase for the reconstruction and not for the downstream tasks including audio-video retrieval.

\subsection{Retrieval with CMMixer}
\textbf{Training.} The synchronization (alignment) between audio and video plays a crucial role in determining how they correspond to each other. Therefore, models that aim to capture this correspondence need to take into account the temporal context in which the audio and video are presented. To accomplish this, we retain the CMMixer encoder and duplicate it to create a two-stream contextual retrieval framework $F=\{f^v, f^a\}$, including video encoder $f^v$ and audio encoder $f^a$ for visual and auditory signal, and initialize them with the same pre-trained weights. We use this synchronization as a form of self-supervision, training our model to use separated modality data $I^v$ and $I^m$ as input to get encoded output $P^v$ and $P^m$ respectively to measure the similarity of corresponding music and video segments. The objective function measures the similarity between the video and audio representation segments, maximizes the mutual information between positive pairs, and minimizes the mutual information between negative pairs. This is achieved through the InfoNCE contrastive loss function, which can be formulated as Eq.\ref{eq2}:
\begin{equation}
    \mathcal{L}= -\sum_{i}^{V} \sum_{l}^{L} \left[ \log{\frac{\exp{\left(s(P_{i,l}^v,P_{i,l}^m)/\tau\right)}}{\sum_{j}^{V} \sum_{l}^{L}\exp{\left(s(P_{i,l}^v,P_{j,l}^m)/\tau\right)}}} \right]
    \label{eq2}
\end{equation}


Where $s(P^v,P^m)$ is the similarity function, $\tau$ is a hyperparameter that we set to $\tau = 0.3$.

\textbf{Inference.} During the inference phase, the proposed model receives a short video segment as input and generates a recommendation for a matching audio track that aligns with both the visual content and contextual information of the video. Meanwhile, it can also recommend a video clip that aligns with a given music track. The audio and video segments are selected from a pool of all available segments in the test dataset using the similarity metric learned during the model's training phase. The model has the ability to match both audio and video in a similar manner. The objective of this process is to effectively align the recommended music and video segments, ensuring a harmonious and coherent relationship between the two modalities.

\begin{figure}[htbp]
    \centering
    \includegraphics[scale=0.4]{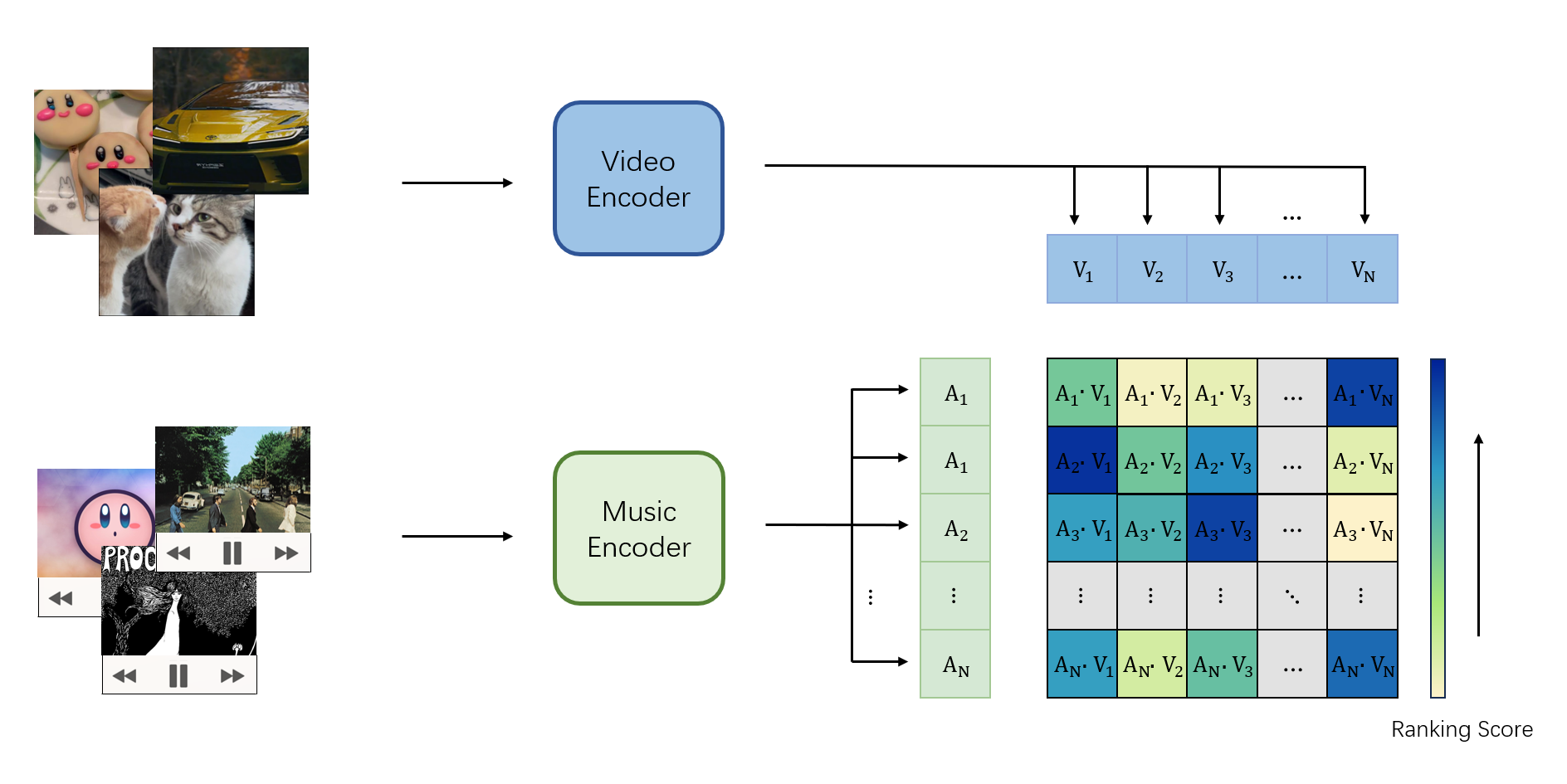}
    \caption{\textbf{Overview of CMMixer Retrieval}. It learns encoders for both audio and images in the context of video-music matching using dual encoder models, which are initialized with the same set of weight parameters and trained with contrastive losses.
}
    \label{fig:mmixer_retrieval}
\end{figure}

\section{EXPERIMENTS}

In this section, we conduct a comprehensive study to evaluate the representation quality of CMMixer using a variety of tasks including audio-video retrieval and other downstream tasks. Our goal is to gain a deeper understanding of CMMixer's performance and potential by answering the following key questions: 

\begin{itemize}
    \item[\textbf{RQ1:}] How does CMMixer compare to the current state-of-the-art methods?
    \item[\textbf{RQ2:}] What is the effect of different mixing methods on fusion performance?
    \item[\textbf{RQ3:}] Does the learned representation incorporate meaningful information from both images and audio?
\end{itemize}

\subsection{Datasets}
\begin{itemize}
    \item \textbf{M2M} includes 10,000 videos that range in length from 10 to 60 seconds. It consists of 2,000 videos with natural sound only, and the remaining 8,000 videos are paired with a specific music clip from a set of 3762 candidate music options. This is done to increase the difficulty of correctly matching the video and audio for the video-music retrieval task and to improve the performance of the model. We use M2M for retrieval tasks and attribute conditioning~\cite{suris2022s}.
    \item \textbf{YouTube-8M} is a large-scale video dataset, which includes more than 7 million videos with 4716 classes labeled by the annotation system. Each video contains audio and visual modalities. Based on the visual information, videos are divided into 24 topics, such as sports, games, arts \& entertainment, etc. In the experiments, we use the original videos as input rather than provided features extracted by the pre-trained models. We select approximately 100,000 videos randomly to construct a balanced subset of it and use it for video-music retrieval tasks.
    \item \textbf{Kinetics-400~\cite{carreira2017quo}} is a comprehensive collection of YouTube videos, featuring 400 human-centered action classes. Each class comprises at least 400 video clips of roughly 10 seconds in length. The actions included in the dataset cover a wide range of categories, including interactions between humans and objects (e.g. playing instruments) and interactions between humans (e.g. handshakes). In the experiments, the dataset has been meticulously cleaned, with a 5 \% sample of approximately 9000 videos containing background sound being selected for the action classification. 
    \item \textbf{AudioSet~\cite{gemmeke2017audio}} is a collection of about 2 million short video clips that are used for audio classification. Each clip has labels for 527 different types of audio events, and multiple events can be in a single clip. The dataset is divided into two subsets: a balanced subset with 22,176 clips and an unbalanced subset with 2,042,985 clips. In the experiments, we use balanced training clips and evaluation clips. Results are reported using the accuracy on the evaluation set.
    \item \textbf{Environmental Sound Classification (ESC-50)~\cite{piczak2015esc}} is a collection of 2,000 5-second audio recordings of environmental sounds. The dataset consists of 50 different classes of sounds. The results are reported as accuracy.
\end{itemize}

\subsection{Comparison Methods}
\begin{itemize}
    \item \textbf{CBMVR}~\cite{hong2018cbvmr} is a two-branch neural network considering two losses to constraint semantics similarity and modality-specific characteristics of embedding, which consist of pre-train CNN to extract video feature and manual feature for audio.
    \item \textbf{CMMVR}~\cite{pretet2021cross} constructs a deeper neural network by integrating both video and audio models pre-trained on cross-modal tasks and previous frames to achieve music-video recommendations based on context.
    \item \textbf{CVMVAE}~\cite{yi2021cross} is a hierarchical Bayesian generative model that matches relevant background music to a micro-video by projecting these two multimodal inputs into a shared latent space through cross-generation. It also introduces the text labels as auxiliary information to enhance performance. 
    \item \textbf{MVPt}~\cite{suris2022s} is one of the state-of-the-art video-music recommendation models. It leverages large pre-train models of image and audio and proposes modeling the long-term temporal context of both the video frame and the audio frame by implementing late fusion.

\end{itemize}

\subsection{Experiment Setup}

\textbf{Data formatting.} Based on VideoMAE~\cite{feichtenhofer2022masked} and AudioMAE~\cite{huang2022masked}, we divide the video into $n$ clips (with a default setting of 8) and extract one frame and 2.6 seconds waveform with a 16,000 sampling rate from the middle of each clip. Random resized cropping with a scale range of $[0.5, 1]$ and random horizontal flipping are applied to the video frames. For audio, we use only mel spectrograms and do not apply any augmentations. To reduce the cost of data pre-processing, we apply repeated sampling~\cite{hoffer2020augment} to alleviate the bottleneck caused by large data loading. The raw waveform is converted into Mel-frequency bands using a 25ms Hanning window with a 10ms shift. The resulting images and spectrograms have the dimension of $8 \times 224 \times 244$ in the default setting.

\textbf{Pre-training setup.} Our pre-training task for the CMMixer is reconstructing original data using cross-modal fusion information with an autoencoder. The pre-training setup for CMMixer follows the standard ViT architecture, using a ViT-base encoder and decoder(s). The encoder has 12 layers and the decoder(s) have 4 layers. The mask ratio is set to 0.5. For a single decoder, we use the Adam optimizer with a learning rate of $2.e-4$. For two separate decoders, the learning rate is set at $1.e-4$. Both have a weight decay of 0.05 and a momentum of (0.9, 0.95). We train for 400 epochs with a batch size of 32 and use a warmup of 80 epochs, followed by cosine decay with a minimal learning rate of $1.e-5$.

\textbf{Evaluation setup.} We evaluate the performance of our model using fine-tuning. We apply supervised fine-tuning on AudioSet, ESC-500, and Kinetics-400 datasets for 100 epochs each with a learning rate of $2.e-4$. Additionally, we also perform unsupervised fine-tuning on our M2M dataset, also with a learning rate of $2.e-4$.

\subsection{Evaluation Metrics}
In our retrieval experiments, we evaluate the quality of retrievals under different model designs and training regimes. The models are pre-trained on the M2M dataset and fine-tuned in an unsupervised manner on the same dataset to create the retrieval model by retaining only the encoder. Specifically, we retrieve complete music audio tracks given a query video and vice versa to assess the quality of the retrievals.

In order to evaluate the success of retrieval, we compute the feature distance between a given query track (either visual or musical) and each of the $\textbf{N=2000}$ target candidates in a pool, which have not been seen during the training of the model. This pool of target candidates contains one correct pair, the ground truth. The candidates are then ranked based on the computed feature distance values and evaluated using two different criteria.

As per usual, we introduce Recall@K as one of the evaluation metrics. This metric measures the success of the retrieval by considering the K closest candidates. If the ground truth pair is present among the K closest candidates, then the retrieval is considered successful and the percentage of successful retrievals in the test set is reported. Another evaluation metric is Median Rank \cite{suris2022s}, which determines the position of the ground truth pair in the sorted list of candidates. The median of the position values across the test set is then reported as the final result.

\begin{table*}[htbp]
\LARGE
\centering
\caption{\textbf{Quantitative comparisons on retrieval classification tasks.} The results showed that the combination of CutMix-Solo yielded the best performance}
\label{table1}
\renewcommand{\arraystretch}{1.05}
\resizebox{1.0\textwidth}{!}{
\setlength{\tabcolsep}{5.2mm}{
\begin{tabular}{c c c c c c c c c c}
\toprule[1.25pt]

\multirow{2}{*}{\textbf{Architecture}} & \multirow{2}{*}{\textbf{Mixer}} & V $\rightarrow$ M & M $\rightarrow$ V & \multicolumn{3}{c}{V $\rightarrow$ M} & \multicolumn{3}{c}{M $\rightarrow$ V}\\

\cmidrule(r){3-3}\cmidrule(r){4-4}\cmidrule(r){5-7}\cmidrule(r){8-10}

\multirow{2}{*}{} & \multirow{2}{*}{} & \multicolumn{2}{c}{Median Rank $\downarrow$} & \multicolumn{1}{c}{R@1$\uparrow$}& R@5 & \multicolumn{1}{c}{R@10}& R@1 & R@5 & R@10\\
\midrule[1.25pt]
\multirow{3}{*}{Solo}&\multicolumn{1}{c}{CutMix}& 6 & 6 &\multicolumn{1}{c}{11.2}&\textbf{34.6}&\multicolumn{1}{c}{\textbf{58.5}}&\textbf{10.9}&33.4&\textbf{56.8}\\
\multirow{3}{*}{}&\multicolumn{1}{c}{Pixel Mixup}& 14&15 &\multicolumn{1}{c}{9.7}&28.3&\multicolumn{1}{c}{53.2}&9.5&27.5&52.1\\
\multirow{3}{*}{}&\multicolumn{1}{c}{Mixup}&22 & 19 &\multicolumn{1}{c}{4.4}&17.3&\multicolumn{1}{c}{30.7}&4.1&16.6&29.4\\
\midrule
\multirow{3}{*}{Duet}&\multicolumn{1}{c}{CutMix}& 5 & 8 &\multicolumn{1}{c}{\textbf{11.4}}&32.8&\multicolumn{1}{c}{56.2}&10.3&\textbf{34.7}&53.8\\
\multirow{3}{*}{}&\multicolumn{1}{c}{Pixel Mixup}& 11 & 12 &\multicolumn{1}{c}{9.3}&28.2&\multicolumn{1}{c}{55.3}&10.3&34.7&53.8\\
\multirow{3}{*}{}&\multicolumn{1}{c}{Mixup}& 25 &28 &\multicolumn{1}{c}{4.8}&18.5&\multicolumn{1}{c}{32.6}&4.3&17.3&31.5\\
\bottomrule[1.25pt]
\end{tabular}
}
}

\label{table_MAP1}
\end{table*}

\subsection{Performance Comparison (RQ1)}

Our work builds upon the latest advancements in cross-modal matching methods for video-music retrieval tasks, including CBMVR \cite{hong2018cbvmr}, CMMVR \cite{pretet2021cross}, CMVAE \cite{yi2021cross}, and MVPt \cite{suris2022s}. In order to ensure a fair comparison and consistent experimental details, some modifications were made to the implementation of these methods. The results are presented in Table.~\ref{table_MAP6}, where our proposed method demonstrates superior performance compared to the others, achieving the best results on the M2M and YouTube-8M metrics.

Among previous methods, CBVMR and CMMVR obtain the worst performance in two datasets. This is because CBVMR only uses a simple neural network to process unsuitable pre-train video features and a few manual features for audio. CMMVR does its effort based on CBVMR while it just uses some pre-train modules to build their framework without considering rationality, which lets to worse performance. The CMVAE has better performance due to introducing text labels for other information to enrich the input data. MVPt gets the best performance among the previous models by using a large pre-train model for both video and audio. This architecture obtains more fine-grained modality features and achieves better performance. For the result in M2M and YouTube-8M, the Recall@1 in YouTube-8M is relatively lower because of only conducting fine-tuning on it. But the trend between them is similar. These previous models often assume a strong semantic correlation between the overall video and audio for the input data and use a relatively simple dual-tower network structure to capture it. However, in practical application scenarios, this assumption is often invalid. For example, there is only a local correlation between visual and audio in many cases so the dual-tower models have difficulty capturing this correlation.

Our model beat all the former methods and achieves significantly higher accuracy, even outperforming the prior state-of-the-art up to $2\times$ and also maintaining the best performance when transferring to YouTube-8M. We propose an optimization method, an intermediate step whereby we fuse modalities $I_v$ and $I_m$ into $I_i$ and use it to reconstruct images that force modality alignment. More specifically, we use frame-level attention to implicitly measure cross-modality sub-element correlation. The attention weights represent the correlation. In image reconstruction, the attention determines what regions, either the same or different modalities, to focus on when generating modality parts. This maps each modality to a shared feature space and learns a modality-common representation.

Our single-tower structure constrains each modality to enter a coordinated space by optimizing similarity between mapped modalities, providing intermediate steps from unpaired audio/video to paired data. Then, we use a double-tower cross-modal retrieval structure to specialize from the joint space provided by a single-tower pre-training model to each modality space, representing each modality separately. This method reduces unpaired-to-paired difficulty, handles hard examples, and improves accuracy.

\begin{table}[htbp]
\footnotesize
\centering
\caption{\textbf{Comparison with other models in video-music retrieval task.} Recall@1 on M2M and YouTube-8M is reported and each top score is highlighted in bold.}
\label{table_MAP6}
\renewcommand{\arraystretch}{1.05}
\setlength{\tabcolsep}{7mm}{
\begin{tabular}{c c c}
\midrule[1.25pt]
\textbf{Model}&\textbf{M2M} & \textbf{YouTube-8M}\\
\hline
CBVMR & 2.73 & 1.61\\
CMMVR & 1.49 & 1.33\\
CVMVAE & 4.05 & 3.27\\
MVPt & 6.22 & 5.69\\
\hline
CMMixer-Solo & \textbf{11.40} & \textbf{10.82}\\
CMMixer-Duet & 8.92 & 8.64\\
\midrule[1.25pt]
\end{tabular}}
\label{table_MAP6}
\end{table}

\subsection{The Effect of Mixing Methods and Architectures (RQ2)}

To explore the cross-modal modeling of our framework, we show how modifying each of the model components contributes to an increase in retrieval performance. We compare the results on six variants for pre-training: two architectures with three subtypes of the mixer.

\begin{figure*}[htbp]
    \centering
    \includegraphics[scale=0.4]{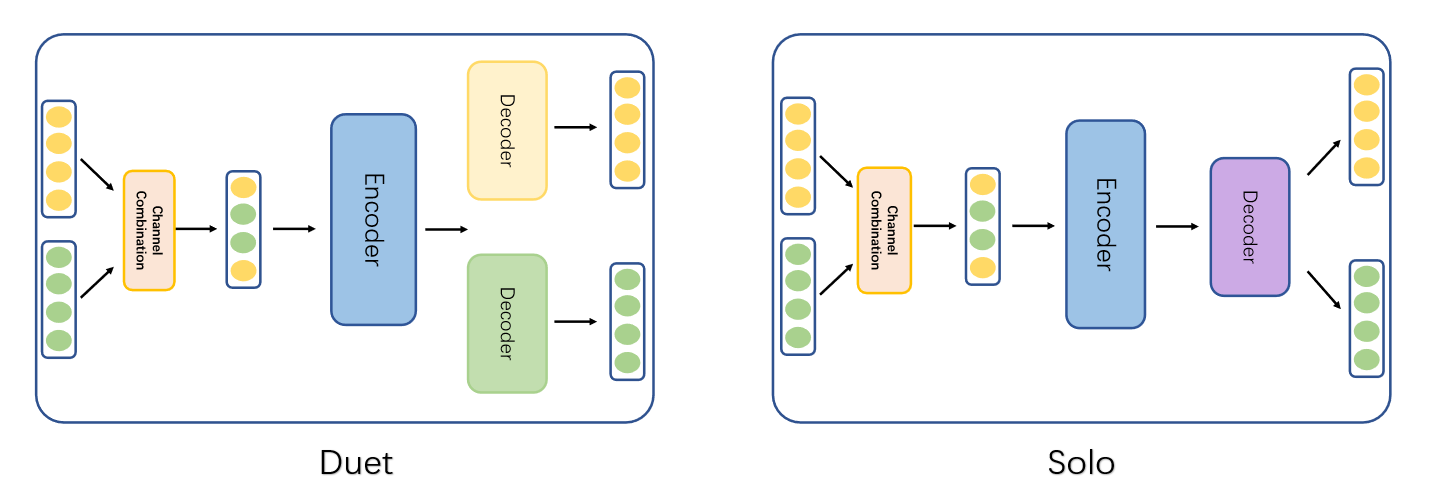}
    \caption{\textbf{Comparison of two types of CMMixer Decoder.}}
    \label{fig:CMMixer_decoder}
\end{figure*}

We employ a different model structure, referred to as \textit{"Duet"}, which features two independent decoders. This architecture is distinct from the one previously mentioned, referred to as \textit{"Solo"}, which employs a single shared decoder. For the Duet structure, we have designed two separate decoders, denoted as $D=\{D^v,D^a\}$, that are responsible for decoding the different representational spaces of the two modalities. The inputs for these two decoders consist of the shared encoded visible patches (p) and the two modalities mask tokens ${m^v,m^a}$. Specifically, $I^v=\{m^v,p\}$ and $I^a=\{m^a,p\}$. The design of these two decoders is similar to the first strategy, including the use of position embeddings and stacks of Transformer blocks in the decoder. However, in the final layer, each decoder has an independent linear projection that outputs the pixel values in a patch for reconstructions $R^v = D(I^v), R^a = D(I^a)$. Note that the MLP of the decoder(s) has access to the same set of base features for each modality and in our model, we match the number of model parameters of single encoder and decode. The detail of our framework is illustrated in Fig.~\ref{fig:CMMixer_decoder} and Table.~\ref{table4}. We summarize the performance of different variants of CMMixer for M2M in Table.~\ref{table_MAP1}, from which we have the following observations.

In the experiments, the CMMixer-Solo with a single decoder has better performance than CMMixer-Deut, indicating that the modality-common representations are essential. In comparison, the representation learning ability of the CMMixer-Duet model was hindered by the inappropriate use of two decoders, forcing the model to learn modality-specific representations that should be learned during downstream task fine-tuning, resulting in a significant difference in generalization compared to the CMMixer-Solo model. 

For the subtype of the CMMixer, we can observe that CutMix consistently outperforms Pixel Mixup and Mixup. This is because Pixel Mixup has a more discrete and fragmented modality fusion view, which results in difficulty for the modal reconstruction by utilizing continuous content. The Mixup seems intuitive for the human being, it introduces discrepancy, makes the model confused with noise, and destroys the topology of information. In contrast, CutMix has a clear boundary, allowing the model to better use semantics for modeling and achieving better results. 

Therefore, for the whole framework, the amalgamation of CMMixer-Solo and CutMix using a distinct cross-modal view and a single autoencoder can generate the best representation for the retrieval task.


\begin{table}[htbp]
\footnotesize
\centering
\caption{\textbf{Comparison of Params and Flops.} Note that the data here represents the model computational cost during pre-training. The model with a single decoder is smaller and faster.}
\label{table4}
\renewcommand{\arraystretch}{1.05}
\setlength{\tabcolsep}{7mm}{
\begin{tabular}{c c c}
\midrule[1.25pt]
\textbf{Type}&\textbf{Params}&\textbf{FLOPs}\\
\hline
CMMixer-Solo & 91.0M & 47.8\\
CMMixer-Duet & 94.4M & 58.5\\
\midrule[1.25pt]
\end{tabular}}
\label{table_MAP4}
\end{table}

\subsection{Transfer Tasks and Datasets (RQ3)}
In order to verify the effectiveness of our proposed pattern in both visual and auditory modalities, and to explore whether our model training strategy enables it to learn more generalized features beyond just matching knowledge, we assessed the performance of our CMMixer model by conducting fine-tuning experiments on several downstream tasks. We compared our results only to those achieved using the CMMixer-Solo-CutMix approach as the baseline. 



\textbf{Action Classification.} Table.~\ref{table3} compares our model to prior state-of-the-art self-supervised frameworks of the action classification task that rely on pre-training. We evaluate our model on action classification task on the Kinetics-400. Results are reported using the top-1 classification accuracy (\%). For a fair comparison, our main benchmark is the model in the middle group. On the one hand, the CMMixer-Solo achieves 0.6\%, 0.2\%, and 0.4\% higher accuracy than BEVT, VideoMAE-B and ST-MAE-B respectively. On the other hand, our CMMixer-Solo surpasses its counterparts with masked video modeling with a smaller pre-training scale. Compared with VideoMAE with 1600-epoch pre-training in Kinetics-400, the CMMixer with 400-epoch pre-training in M2M obtains accuracy improvement.


\begin{table}[htbp]
\footnotesize
\centering
\caption{\textbf{Comparison with other state-of-the-art models on action classification. }We report top-1 accuracy on the validation set. The “PT data” column specifies the pre-training dataset of the models, IN:ImageNet-1k, K600:Kinetics-600, K400:Kinetics-400.}
\label{table3}
\renewcommand{\arraystretch}{1.05}

\begin{tabular}{c c c c c}
\midrule[1.25pt]
\textbf{Model}&\textbf{Backbone}&\textbf{PT Data}&\textbf{K400}&\textbf{Params}\\
\hline
MaskFeat\cite{wei2022masked} & MViTv2-L & K400 & 84.3 & 218M\\
BEVT\cite{wang2022bevt} & Swin-B & IN+K400 & 81.1 & 88M\\
VideoMAE-L\cite{tong2022videomae} & ViT-L & K400 & 85.2 & 305M\\
VideoMAE-B & ViT-B & K400 & 81.5 & 87M\\
ST-MAE-H\cite{feichtenhofer2022masked} & Vit-H & K400+K600 & 86.8 & 632M\\
ST-MAE-B & ViT-B & K400 & 81.3 & 87M\\
\hline
CMMixer-Solo & ViT-B & M2M & 81.7 & 87M\\
CMMixer-Duet & ViT-B & M2M & 78.3 & 87M\\
\midrule[1.25pt]
\end{tabular}

\label{table_MAP3}
\end{table}

\textbf{Audio classification.} Table.~\ref{table2} compares our model to prior state-of-the-art self-supervised frameworks of the action classification task that rely on pre-training. We evaluate our model on the audio classification task on the AudioSet and ESC-50. Results are reported using the top-1 classification accuracy (\%). Fine-tuning on AudioSet-20K, achieving 34.8\%, CMMixer-Solo significantly outperforms concurrent MAE-AST \cite{baade2022mae}, which trained with an additional 1,000 hours of speech in Librispeech while we only fine-tune without off-domain pre-training. However, we observe that we are still a little short of the best model AudioMAE, which is both pre-trained and fine-tuned on AudioSet and has a larger size of the model. For the task on ESC-50, we report the same higher accuracy, 91.5\%, to AST. This means that we have achieved the standard performance in such a small but meaningful dataset. The results are shown in Table.~\ref{table_MAP2}.

\begin{table}[htbp]
\footnotesize
\centering
\caption{\textbf{Comparison with other state-of-the-art models on audio and speech classification
tasks.}We report mAP for AS and accuracy (\%) for ESC. The “PT data” column specifies the pre-training dataset of the models, AS:AudioSet, LS:LibriSpeech, and IN:ImageNet.}
\label{table2}
\renewcommand{\arraystretch}{1.05}
\setlength{\tabcolsep}{1.61mm}{
\begin{tabular}{c c c c c c }
\midrule[1.25pt]
\textbf{Model}&\textbf{Backbone}&\textbf{PT Data}&\textbf{AS-20K}&\textbf{ESC-50}&\textbf{Params}\\
\hline
AST\cite{baade2022mae}&ViT-B&AS+LS&30.6&90&86M\\
Audio-MAE\cite{huang2022masked}&ViT-B&AS&\textbf{37.1}&\textbf{94.1}&86M\\
\hline
CMMixer-Solo&ViT-B&M2M&34.8&91.5&86M\\
CMMixer-Duet&ViT-B&M2M&30.5&90.2&86M\\
\midrule[1.25pt]
\end{tabular}}

\label{table_MAP2}
\end{table}

\textbf{Discussion.} However, despite not surpassing state-of-the-art models when transferred to other datasets, we still found advantages of our model. Compared to other visual and audio models, our model achieved better performance with the same or fewer data samples under equivalent parameter settings. This indicates that using the CMMixer method allowed us to achieve a stronger utilization of training samples and higher efficiency, which is a trade-off between computational cost and accuracy, shown in Table.~\ref{table5}. Additionally, thanks to the pre-training of our modal fusion based on the masked modeling method, our model learned a good joint representation that can be easily fine-tuned on different datasets to achieve good results.

In summary, our model primarily achieves the generalization of multimodal joint representation by constructing a unique pre-training method, utilizing CMMixer for modality fusion, and implicitly aligning modalities through an encoder-decoder structure with an attention mechanism. This enables our model to capture the interdependence of visual and audio information and capture the long-range temporal context, ultimately leading to superior downstream task performance when used as an intermediate representation. 

\begin{table}[htbp]
\footnotesize
\centering
\caption{\textbf{Comparison with other models on pre-training Dataset scale.}}
\label{table5}
\renewcommand{\arraystretch}{1.05}
\setlength{\tabcolsep}{7mm}{
\begin{tabular}{c c c}
\midrule[1.25pt]
\textbf{Model}&\textbf{PT Data}&\textbf{Dataset scale}\\
\hline
ST-MAE & K600 & 9×\\
MAE & K400 & 5×\\
AST & AS+LS & 55×\\
Audio-MAE & AS & 49×\\
\hline
CMMixer & M2M & 1×\\
\midrule[1.25pt]
\end{tabular}}
\label{table_MAP5}
\end{table}

\section{ANANLYSIS}

\begin{figure*}
  \centering
  \includegraphics[scale=0.0593]{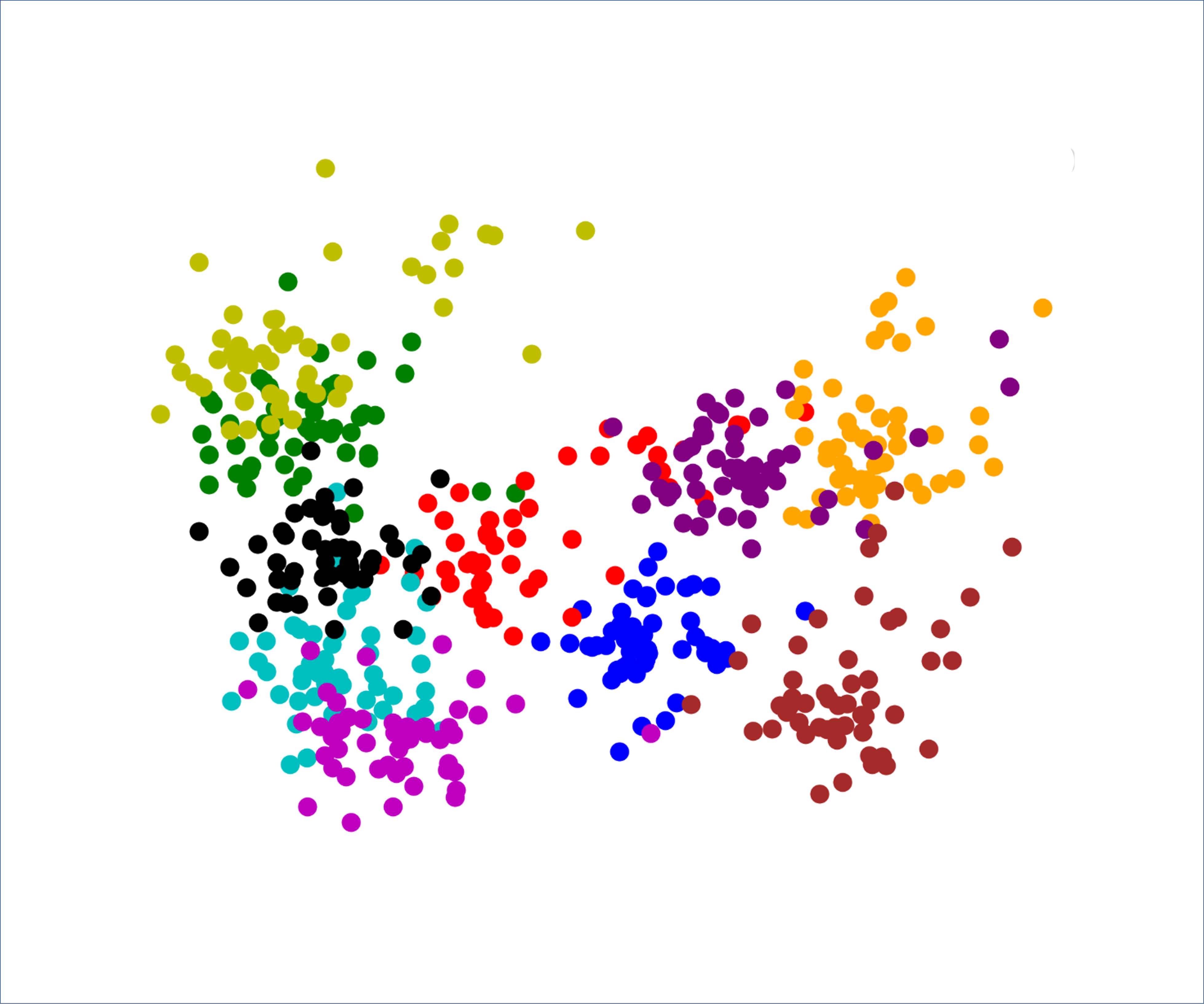}
  \hspace{0.77in}
  \includegraphics[scale=0.0593]{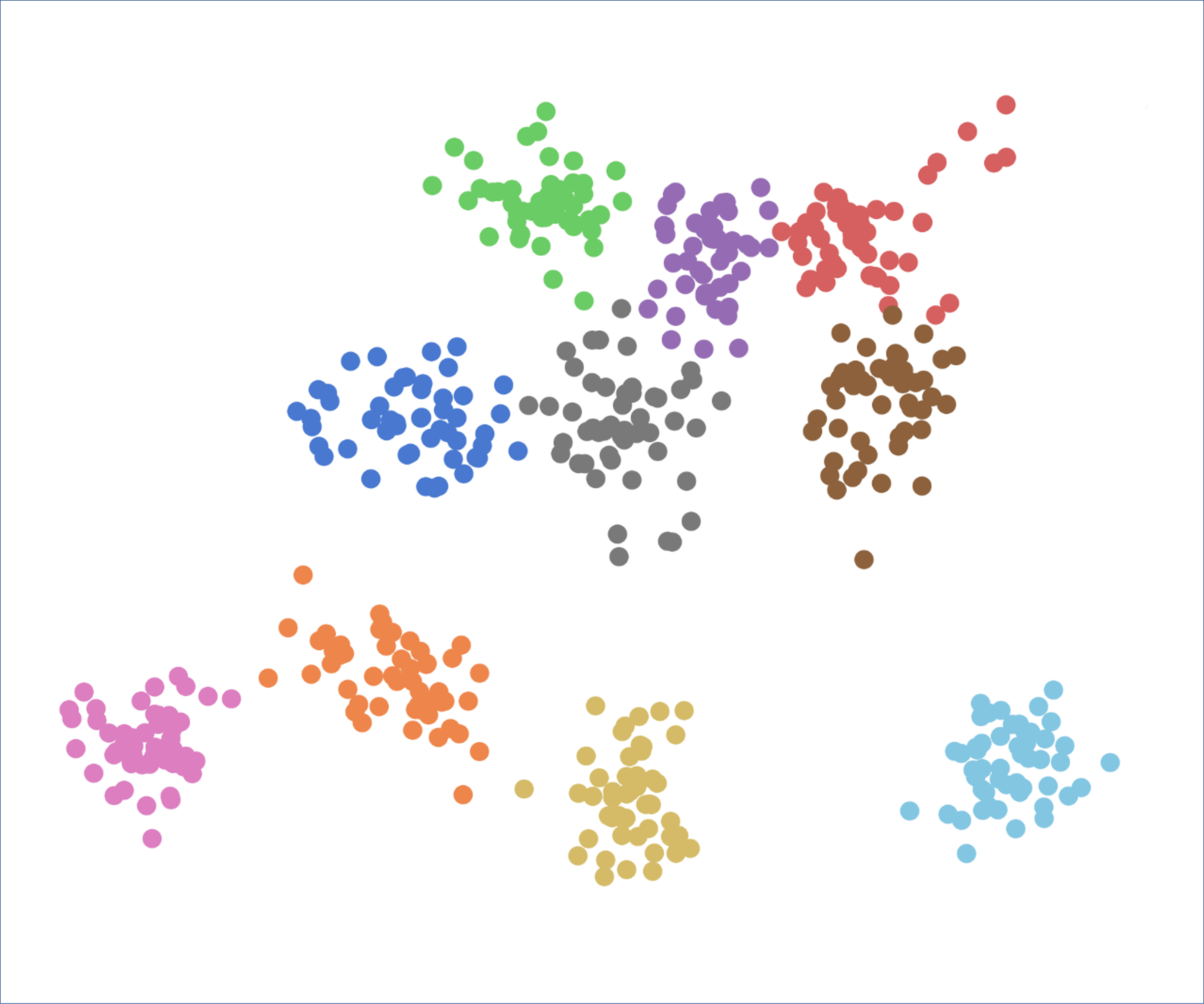}
  \caption{\textbf{t-SNE visualization for learned representations.} Left is music segments and the right is action videos. The points from the same entity have the same color.}
\label{fig:dot}
\end{figure*}

\subsection{Clustering analysis of representation}
We evaluated the learned representations of CMMixer using t-SNE \cite{van2008visualizing} visualizations in two parts – music segments in M2M retrieval and actions in Kinetics-400. For the music segments, we analyzed it based on songs, where we selected videos containing 10 different songs (with the same song but different clips); For the actions, we analyzed it based on visual content, where we selected 10 categories (selected and roughly classified by manual screening). For both action and audio classification, as shown in Fig.~\ref{fig:dot}, we selected 10 categories. Our observations showed that: 
1) The latent representations of the same song or content category are clustered into one cluster; 
2) The latent representations of both visual and audio can be classified into 10 different categories with reasonable accuracy. These observations indicate that the representations learned by CMMixer are versatile and have fair discriminability.

\begin{figure}[htbp]
    \centering
    \includegraphics[scale=0.5]{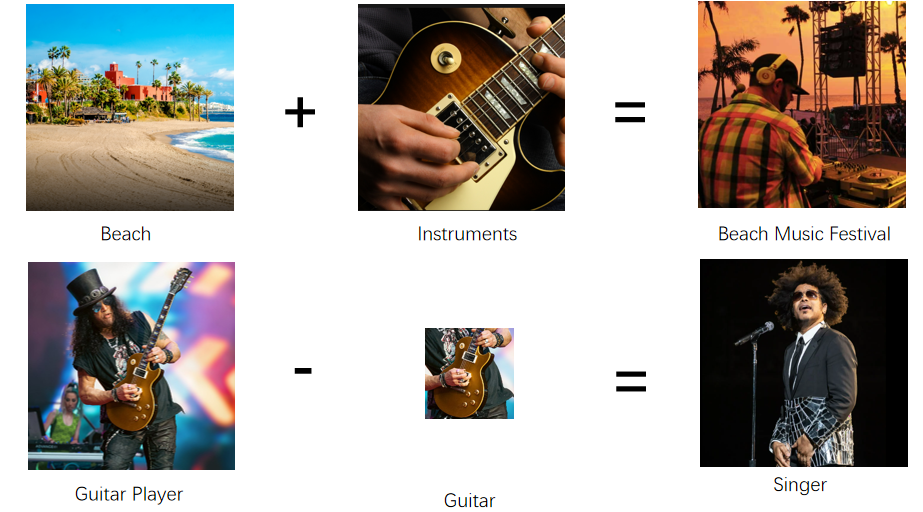}
    \caption{\textbf{Zero-shot cross modality retrieval with flexible queries}. Our model enables zero-shot cross-modal search of flexible queries without any annotated data. By adding or subtracting a text query embedding from an image query embedding, we can retrieve relevant information using cosine similarity}
    \label{fig:0}
\end{figure}

\subsection{Zero-shot cross-modality retrieval with flexible queries}
In the previous description of the retrieval task, we need to input one modality to the model in order to find a suitable one in the resource pool for specific information. However, given that our model captures a wide range of audio-visual representations and possesses strong versatility, we propose to use its representations to condition the retrieval process.

In this process, we have a modality $M$ that represents specific information and to enhance or alter its attributes, we add another query embedding $M_a$ with the same or even another modality as adjuvant information to the original query embedding $M$. We can calculate the representation $M_{transfer}$ in the one single model because of data format consistency: $M_{transfer} = M  +M_a$. If we instead want to remove the attribute, we just use a subtraction: $M_{transfer} = M - M_a$ by padding zero or special tokens if necessary. This approach has various applications, such as video editing, music or video style transfer searching, as demonstrated in Fig.~\ref{fig:0}. For example, if we have a video clip of a beach scene, we can retrieve a similar video clip of a beach music festival by combining it with an audio segment of a pop band ensemble. This new information, $M_{transfer}$, can then be used as an input to retrieve the desired video clip.

\begin{figure*}[t]
  \centering
  \includegraphics[scale=0.3]{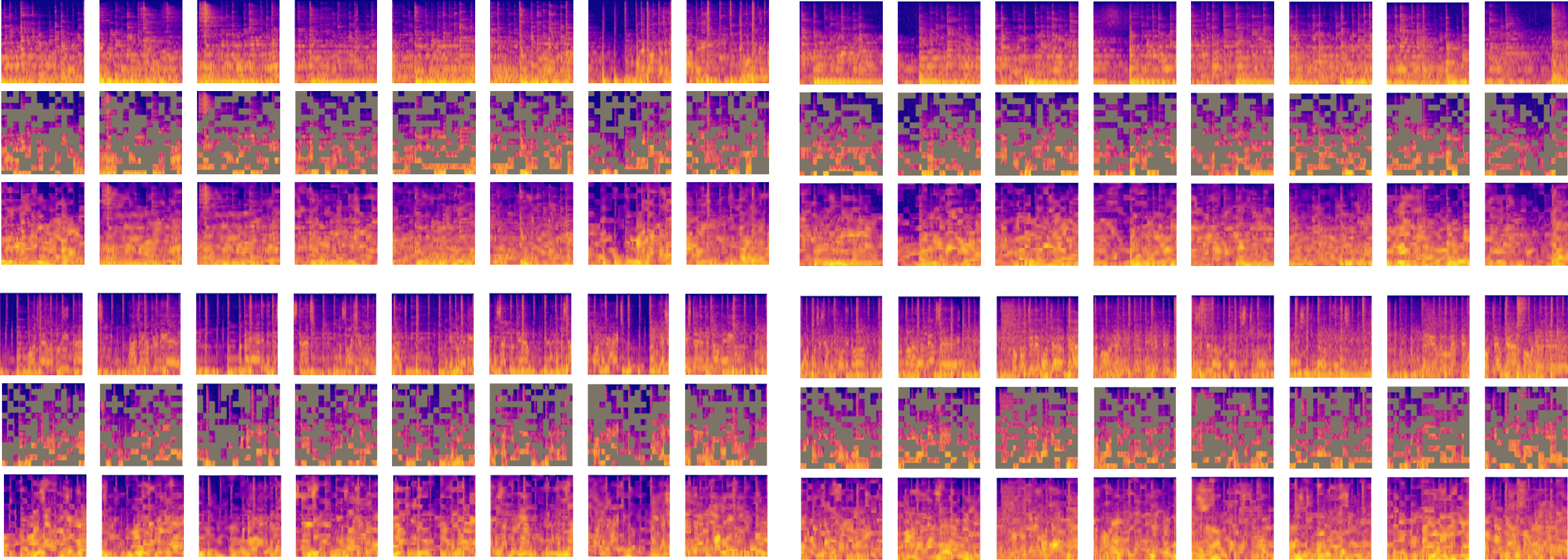}
  \caption{\textbf{Visualization of spectrograms reconstruction.} The examples are reconstructed from the M2M \emph{eval} set using a pre-trained model with a mask ratio of 0.5.}
\label{fig:pic1}
\end{figure*}

\subsection{Reconstruction visualization}
\begin{figure*}[t]
  \centering
  \includegraphics[scale=0.3]{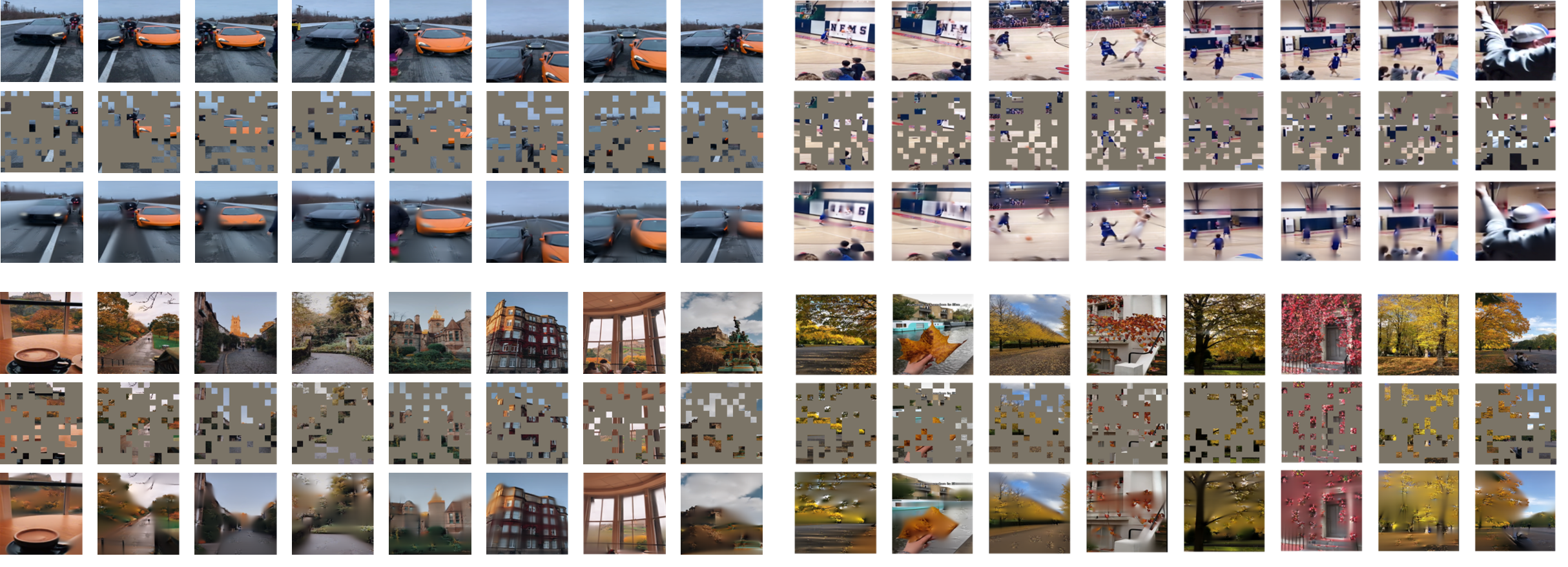}
  \caption{\textbf{Visualization of visual images reconstruction.} The examples are reconstructed from the M2M \emph{eval} set using a pre-trained model with a mask ratio of 0.5.}
\label{fig:pic}
\end{figure*}

We are interested in the reconstruction quality of pre-trained CMMixer. We randomly sample examples from M2M and Kinetics and show the results in Fig.~\ref{fig:pic1} and Fig.~\ref{fig:pic}. In each reconstructed image, we include original unmasked tokens for better visual quality. We observe that our model infers holistic reconstructions across M2M and Kinetics datasets, indicating it has learned numerous concepts.

\section{Conclusion and Future Work}
We present CMMixer, a new multi-modal self-supervised approach for the video-audio retrieval task, using visual and acoustical information to train a reconstruction network as a pre-trained model. We experimented with two model architectures and a channel combination operation to integrate channel information, enabling our method to handle single or multiple modalities with a single-stream encoder efficiently. The results of our evaluations on retrieval and cross-modality matching tasks surpass current state-of-the-art methods. Moreover, our model's versatility is demonstrated by its excellent balance of performance on a range of downstream datasets, even in diverse tasks.

For future work, we hope that we can build on these findings and successes and conduct more experiments such as pre-training in large general dataset video with audio to explore the ability and capacity of our framework to construct a scalable multi-domain strategy for self-supervised pre-training.

\section*{Acknowledgment}
This work is partially supported by the National Natural Science Foundation of China under Grants No. 62032013 and No. 61972078, and the Fundamental Research Funds for the Central Universities under Grants No. N2217004 and No. N2317002.

\bibliography{reference}

\begin{thebibliography}{10}
\expandafter\ifx\csname url\endcsname\relax
  \def\url#1{\texttt{#1}}\fi
\expandafter\ifx\csname urlprefix\endcsname\relax\def\urlprefix{URL }\fi
\expandafter\ifx\csname href\endcsname\relax
  \def\href#1#2{#2} \def\path#1{#1}\fi

\bibitem{pretet2021cross}
L.~Pr{\'e}tet, G.~Richard, G.~Peeters, Cross-modal music-video recommendation:
  A study of design choices, in: 2021 International Joint Conference on Neural
  Networks (IJCNN), IEEE, 2021, pp. 1--9.

\bibitem{yi2021cross}
J.~Yi, Y.~Zhu, J.~Xie, Z.~Chen, Cross-modal variational auto-encoder for
  content-based micro-video background music recommendation, IEEE Transactions
  on Multimedia.

\bibitem{suris2022s}
D.~Sur{\'\i}s, C.~Vondrick, B.~Russell, J.~Salamon, It's time for artistic
  correspondence in music and video, in: Proceedings of the IEEE/CVF Conference
  on Computer Vision and Pattern Recognition, 2022, pp. 10564--10574.

\bibitem{chen2020simple}
T.~Chen, S.~Kornblith, M.~Norouzi, G.~Hinton, A simple framework for
  contrastive learning of visual representations, in: International conference
  on machine learning, PMLR, 2020, pp. 1597--1607.

\bibitem{cheng2022vista}
M.~Cheng, Y.~Sun, L.~Wang, X.~Zhu, K.~Yao, J.~Chen, G.~Song, J.~Han, J.~Liu,
  E.~Ding, et~al., Vista: vision and scene text aggregation for cross-modal
  retrieval, in: Proceedings of the IEEE/CVF Conference on Computer Vision and
  Pattern Recognition, 2022, pp. 5184--5193.

\bibitem{wang2022vlmixer}
T.~Wang, W.~Jiang, Z.~Lu, F.~Zheng, R.~Cheng, C.~Yin, P.~Luo, Vlmixer: Unpaired
  vision-language pre-training via cross-modal cutmix, in: International
  Conference on Machine Learning, PMLR, 2022, pp. 22680--22690.

\bibitem{chen2022improving}
F.~Chen, X.~Chen, S.~Xu, B.~Xu, Improving cross-modal understanding in visual
  dialog via contrastive learning, in: ICASSP 2022-2022 IEEE International
  Conference on Acoustics, Speech and Signal Processing (ICASSP), IEEE, 2022,
  pp. 7937--7941.

\bibitem{wang2022eclip}
J.~Wang, H.~Wang, W.~Wu, J.~Deng, Y.~Lu, X.~Guo, D.~Zhang, Eclip: Efficient
  contrastive language-image pretraining via ensemble confidence learning and
  masked language modeling, in: First Workshop on Pre-training: Perspectives,
  Pitfalls, and Paths Forward at ICML 2022.

\bibitem{devlin2018bert}
J.~Devlin, M.-W. Chang, K.~Lee, K.~Toutanova, Bert: Pre-training of deep
  bidirectional transformers for language understanding, arXiv preprint
  arXiv:1810.04805.

\bibitem{he2022masked}
K.~He, X.~Chen, S.~Xie, Y.~Li, P.~Doll{\'a}r, R.~Girshick, Masked autoencoders
  are scalable vision learners, in: Proceedings of the IEEE/CVF Conference on
  Computer Vision and Pattern Recognition, 2022, pp. 16000--16009.

\bibitem{vaswani2017attention}
A.~Vaswani, N.~Shazeer, N.~Parmar, J.~Uszkoreit, L.~Jones, A.~N. Gomez,
  {\L}.~Kaiser, I.~Polosukhin, Attention is all you need, Advances in neural
  information processing systems 30.

\bibitem{bao2021beit}
H.~Bao, L.~Dong, S.~Piao, F.~Wei, Beit: Bert pre-training of image
  transformers, arXiv preprint arXiv:2106.08254.

\bibitem{xie2022simmim}
Z.~Xie, Z.~Zhang, Y.~Cao, Y.~Lin, J.~Bao, Z.~Yao, Q.~Dai, H.~Hu, Simmim: A
  simple framework for masked image modeling, in: Proceedings of the IEEE/CVF
  Conference on Computer Vision and Pattern Recognition, 2022, pp. 9653--9663.

\bibitem{liu2022mixmim}
J.~Liu, X.~Huang, Y.~Liu, H.~Li, Mixmim: Mixed and masked image modeling for
  efficient visual representation learning, arXiv preprint arXiv:2205.13137.

\bibitem{zhang2022mae}
K.~Zhang, Z.~Shen, i-mae: Are latent representations in masked autoencoders
  linearly separable?, arXiv preprint arXiv:2210.11470.

\bibitem{agrawal2015learning}
P.~Agrawal, J.~Carreira, J.~Malik, Learning to see by moving, in: Proceedings
  of the IEEE international conference on computer vision, 2015, pp. 37--45.

\bibitem{wang2015unsupervised}
X.~Wang, A.~Gupta, Unsupervised learning of visual representations using
  videos, in: Proceedings of the IEEE international conference on computer
  vision, 2015, pp. 2794--2802.

\bibitem{pathak2017learning}
D.~Pathak, R.~Girshick, P.~Doll{\'a}r, T.~Darrell, B.~Hariharan, Learning
  features by watching objects move, in: Proceedings of the IEEE conference on
  computer vision and pattern recognition, 2017, pp. 2701--2710.

\bibitem{wang2019learning}
X.~Wang, A.~Jabri, A.~A. Efros, Learning correspondence from the
  cycle-consistency of time, in: Proceedings of the IEEE/CVF Conference on
  Computer Vision and Pattern Recognition, 2019, pp. 2566--2576.

\bibitem{misra2016shuffle}
I.~Misra, C.~L. Zitnick, M.~Hebert, Shuffle and learn: unsupervised learning
  using temporal order verification, in: Computer Vision--ECCV 2016: 14th
  European Conference, Amsterdam, The Netherlands, October 11--14, 2016,
  Proceedings, Part I 14, Springer, 2016, pp. 527--544.

\bibitem{fernando2017self}
B.~Fernando, H.~Bilen, E.~Gavves, S.~Gould, Self-supervised video
  representation learning with odd-one-out networks, in: Proceedings of the
  IEEE conference on computer vision and pattern recognition, 2017, pp.
  3636--3645.

\bibitem{lee2017unsupervised}
H.-Y. Lee, J.-B. Huang, M.~Singh, M.-H. Yang, Unsupervised representation
  learning by sorting sequences, in: Proceedings of the IEEE international
  conference on computer vision, 2017, pp. 667--676.

\bibitem{wei2018learning}
D.~Wei, J.~J. Lim, A.~Zisserman, W.~T. Freeman, Learning and using the arrow of
  time, in: Proceedings of the IEEE conference on computer vision and pattern
  recognition, 2018, pp. 8052--8060.

\bibitem{xu2019self}
D.~Xu, J.~Xiao, Z.~Zhao, J.~Shao, D.~Xie, Y.~Zhuang, Self-supervised
  spatiotemporal learning via video clip order prediction, in: Proceedings of
  the IEEE/CVF Conference on Computer Vision and Pattern Recognition, 2019, pp.
  10334--10343.

\bibitem{walker2016uncertain}
J.~Walker, C.~Doersch, A.~Gupta, M.~Hebert, An uncertain future: Forecasting
  from static images using variational autoencoders, in: Computer Vision--ECCV
  2016: 14th European Conference, Amsterdam, The Netherlands, October 11--14,
  2016, Proceedings, Part VII 14, Springer, 2016, pp. 835--851.

\bibitem{vondrick2016anticipating}
C.~Vondrick, H.~Pirsiavash, A.~Torralba, Anticipating visual representations
  from unlabeled video, in: Proceedings of the IEEE conference on computer
  vision and pattern recognition, 2016, pp. 98--106.

\bibitem{mathieu2015deep}
M.~Mathieu, C.~Couprie, Y.~LeCun, Deep multi-scale video prediction beyond mean
  square error, arXiv preprint arXiv:1511.05440.

\bibitem{lotter2016deep}
W.~Lotter, G.~Kreiman, D.~Cox, Deep predictive coding networks for video
  prediction and unsupervised learning, arXiv preprint arXiv:1605.08104.

\bibitem{vondrick2018tracking}
C.~Vondrick, A.~Shrivastava, A.~Fathi, S.~Guadarrama, K.~Murphy, Tracking
  emerges by colorizing videos, in: Proceedings of the European conference on
  computer vision (ECCV), 2018, pp. 391--408.

\bibitem{diba2019dynamonet}
A.~Diba, V.~Sharma, L.~V. Gool, R.~Stiefelhagen, Dynamonet: Dynamic action and
  motion network, in: Proceedings of the IEEE/CVF International Conference on
  Computer Vision, 2019, pp. 6192--6201.

\bibitem{huang2022masked}
P.-Y. Huang, H.~Xu, J.~Li, A.~Baevski, M.~Auli, W.~Galuba, F.~Metze,
  C.~Feichtenhofer, Masked autoencoders that listen, Advances in Neural
  Information Processing Systems 35 (2022) 28708--28720.

\bibitem{tong2022videomae}
Z.~Tong, Y.~Song, J.~Wang, L.~Wang, Videomae: Masked autoencoders are
  data-efficient learners for self-supervised video pre-training, arXiv
  preprint arXiv:2203.12602.

\bibitem{feichtenhofer2022masked}
C.~Feichtenhofer, Y.~Li, K.~He, et~al., Masked autoencoders as spatiotemporal
  learners, Advances in neural information processing systems 35 (2022)
  35946--35958.

\bibitem{geng2022multimodal}
X.~Geng, H.~Liu, L.~Lee, D.~Schuurams, S.~Levine, P.~Abbeel, Multimodal masked
  autoencoders learn transferable representations, arXiv preprint
  arXiv:2205.14204.

\bibitem{peng2022beit}
Z.~Peng, L.~Dong, H.~Bao, Q.~Ye, F.~Wei, Beit v2: Masked image modeling with
  vector-quantized visual tokenizers, arXiv preprint arXiv:2208.06366.

\bibitem{wang2022image}
W.~Wang, H.~Bao, L.~Dong, J.~Bjorck, Z.~Peng, Q.~Liu, K.~Aggarwal, O.~K.
  Mohammed, S.~Singhal, S.~Som, et~al., Image as a foreign language: Beit
  pretraining for all vision and vision-language tasks, arXiv preprint
  arXiv:2208.10442.

\bibitem{he2020momentum}
K.~He, H.~Fan, Y.~Wu, S.~Xie, R.~Girshick, Momentum contrast for unsupervised
  visual representation learning, in: Proceedings of the IEEE/CVF conference on
  computer vision and pattern recognition, 2020, pp. 9729--9738.

\bibitem{oord2018representation}
A.~v.~d. Oord, Y.~Li, O.~Vinyals, Representation learning with contrastive
  predictive coding, arXiv preprint arXiv:1807.03748.

\bibitem{grill2020bootstrap}
J.-B. Grill, F.~Strub, F.~Altch{\'e}, C.~Tallec, P.~Richemond, E.~Buchatskaya,
  C.~Doersch, B.~Avila~Pires, Z.~Guo, M.~Gheshlaghi~Azar, et~al., Bootstrap
  your own latent-a new approach to self-supervised learning, Advances in
  neural information processing systems 33 (2020) 21271--21284.

\bibitem{wu2018unsupervised}
Z.~Wu, Y.~Xiong, S.~X. Yu, D.~Lin, Unsupervised feature learning via
  non-parametric instance discrimination, in: Proceedings of the IEEE
  conference on computer vision and pattern recognition, 2018, pp. 3733--3742.

\bibitem{radford2021learning}
A.~Radford, J.~W. Kim, C.~Hallacy, A.~Ramesh, G.~Goh, S.~Agarwal, G.~Sastry,
  A.~Askell, P.~Mishkin, J.~Clark, et~al., Learning transferable visual models
  from natural language supervision, in: International conference on machine
  learning, PMLR, 2021, pp. 8748--8763.

\bibitem{jia2021scaling}
C.~Jia, Y.~Yang, Y.~Xia, Y.-T. Chen, Z.~Parekh, H.~Pham, Q.~Le, Y.-H. Sung,
  Z.~Li, T.~Duerig, Scaling up visual and vision-language representation
  learning with noisy text supervision, in: International Conference on Machine
  Learning, PMLR, 2021, pp. 4904--4916.

\bibitem{gao2021simcse}
T.~Gao, X.~Yao, D.~Chen, Simcse: Simple contrastive learning of sentence
  embeddings, arXiv preprint arXiv:2104.08821.

\bibitem{kuo2013background}
F.-F. Kuo, M.-K. Shan, S.-Y. Lee, Background music recommendation for video
  based on multimodal latent semantic analysis, in: 2013 IEEE International
  Conference on Multimedia and Expo (ICME), IEEE, 2013, pp. 1--6.

\bibitem{shah2014advisor}
R.~R. Shah, Y.~Yu, R.~Zimmermann, Advisor: Personalized video soundtrack
  recommendation by late fusion with heuristic rankings, in: Proceedings of the
  22nd ACM international conference on Multimedia, 2014, pp. 607--616.

\bibitem{li2019query}
B.~Li, A.~Kumar, Query by video: Cross-modal music retrieval., in: ISMIR, 2019,
  pp. 604--611.

\bibitem{zeng2018audio}
D.~Zeng, Y.~Yu, K.~Oyama, Audio-visual embedding for cross-modal music video
  retrieval through supervised deep cca, in: 2018 IEEE International Symposium
  on Multimedia (ISM), IEEE, 2018, pp. 143--150.

\bibitem{gan2020foley}
C.~Gan, D.~Huang, P.~Chen, J.~B. Tenenbaum, A.~Torralba, Foley music: Learning
  to generate music from videos, in: Computer Vision--ECCV 2020: 16th European
  Conference, Glasgow, UK, August 23--28, 2020, Proceedings, Part XI 16,
  Springer, 2020, pp. 758--775.

\bibitem{su2020audeo}
K.~Su, X.~Liu, E.~Shlizerman, Audeo: Audio generation for a silent performance
  video, Advances in Neural Information Processing Systems 33 (2020)
  3325--3337.

\bibitem{di2021video}
S.~Di, Z.~Jiang, S.~Liu, Z.~Wang, L.~Zhu, Z.~He, H.~Liu, S.~Yan, Video
  background music generation with controllable music transformer, in:
  Proceedings of the 29th ACM International Conference on Multimedia, 2021, pp.
  2037--2045.

\bibitem{abu2016youtube}
S.~Abu-El-Haija, N.~Kothari, J.~Lee, P.~Natsev, G.~Toderici, B.~Varadarajan,
  S.~Vijayanarasimhan, Youtube-8m: A large-scale video classification
  benchmark, arXiv preprint arXiv:1609.08675.

\bibitem{carreira2017quo}
J.~Carreira, A.~Zisserman, Quo vadis, action recognition? a new model and the
  kinetics dataset, in: proceedings of the IEEE Conference on Computer Vision
  and Pattern Recognition, 2017, pp. 6299--6308.

\bibitem{gemmeke2017audio}
J.~F. Gemmeke, D.~P. Ellis, D.~Freedman, A.~Jansen, W.~Lawrence, R.~C. Moore,
  M.~Plakal, M.~Ritter, Audio set: An ontology and human-labeled dataset for
  audio events, in: 2017 IEEE international conference on acoustics, speech and
  signal processing (ICASSP), IEEE, 2017, pp. 776--780.

\bibitem{piczak2015esc}
K.~J. Piczak, Esc: Dataset for environmental sound classification, in:
  Proceedings of the 23rd ACM international conference on Multimedia, 2015, pp.
  1015--1018.

\bibitem{hong2018cbvmr}
S.~Hong, W.~Im, H.~S. Yang, Cbvmr: content-based video-music retrieval using
  soft intra-modal structure constraint, in: Proceedings of the 2018 ACM on
  international conference on multimedia retrieval, 2018, pp. 353--361.

\bibitem{hoffer2020augment}
E.~Hoffer, T.~Ben-Nun, I.~Hubara, N.~Giladi, T.~Hoefler, D.~Soudry, Augment
  your batch: Improving generalization through instance repetition, in:
  Proceedings of the IEEE/CVF Conference on Computer Vision and Pattern
  Recognition, 2020, pp. 8129--8138.

\bibitem{wei2022masked}
C.~Wei, H.~Fan, S.~Xie, C.-Y. Wu, A.~Yuille, C.~Feichtenhofer, Masked feature
  prediction for self-supervised visual pre-training, in: Proceedings of the
  IEEE/CVF Conference on Computer Vision and Pattern Recognition, 2022, pp.
  14668--14678.

\bibitem{wang2022bevt}
R.~Wang, D.~Chen, Z.~Wu, Y.~Chen, X.~Dai, M.~Liu, Y.-G. Jiang, L.~Zhou,
  L.~Yuan, Bevt: Bert pretraining of video transformers, in: Proceedings of the
  IEEE/CVF conference on computer vision and pattern recognition, 2022, pp.
  14733--14743.

\bibitem{baade2022mae}
A.~Baade, P.~Peng, D.~Harwath, Mae-ast: Masked autoencoding audio spectrogram
  transformer, arXiv preprint arXiv:2203.16691.

\bibitem{van2008visualizing}
L.~Van~der Maaten, G.~Hinton, Visualizing data using t-sne., Journal of machine
  learning research 9~(11).

\end{thebibliography}

\end{document}